\documentclass[12pt]{article}%
\usepackage{amsmath}
\usepackage{amsfonts}
\usepackage{amssymb}
\usepackage{graphicx}%
\begin{document}
\begin{titlepage}
\begin{center}
\renewcommand{\thefootnote}{\fnsymbol{footnote}}
{\Large\bf Gravity in Complex Hermitian Space-Time} \vskip20mm
{\large\bf{Ali H. Chamseddine \footnote{email:
chams@aub.edu.lb}\footnote{Published in "Einstein in Alexandria, The
Scientific Symposium",
Editor Edward Witten, Publisher Bibilotheca Alexandrina, pages 39-53 (2006) }}}\\
\renewcommand{\thefootnote}{\arabic{footnote}}
\vskip2cm {\it Physics Department, American University of Beirut,
Lebanon.\\}
\end{center}
\begin{center}
{\bf Abstract}
\end{center}
A generalized theory unifying gravity with electromagnetism was
proposed by Einstein in 1945. He considered a Hermitian metric on a
real space-time. In this work we review Einstein's idea and
generalize it further to consider gravity in a complex Hermitian space-time.
\end{titlepage}

In the year 1945, Albert Einstein \cite{Ein1}, \cite{Einstein} attempted to
establish a unified field theory by generalizing the relativistic theory of
gravitation. At that time it was thought that the only\ fundamental forces in
nature were gravitation and electromagnetism. Einstein proposed to use a
Hermitian metric whose real part is symmetric and describes the gravitational
field while the imaginary part is antisymmetric and corresponds to the Maxwell
field strengths. The Hermitian symmetry of the metric $g_{\mu\nu}$ is given by%
\[
g_{\mu\nu}\left(  x\right)  =\overline{g_{\nu\mu}\left(  x\right)  },
\]
where
\[
g_{\mu\nu}\left(  x\right)  =G_{\mu\nu}\left(  x\right)  +iB_{\mu\nu}\left(
x\right)  ,
\]
so that $G_{\mu\nu}\left(  x\right)  =G_{\nu\mu}\left(  x\right)  $ and
$B_{\mu\nu}\left(  x\right)  =-B_{\nu\mu}\left(  x\right)  .$ However, the
space-time manifold remains real. The connection $\Gamma_{\mu\nu}^{\rho}$ on
the manifold is not symmetric, and is also not unique. A natural choice,
adopted by Einstein, is to impose the hermiticity condition on the connection
so that $\Gamma_{\nu\mu}^{\rho}=\overline{\Gamma_{\mu\nu}^{\rho}},$ which
implies that its antisymmetric part is imaginary. The connection $\Gamma$ is
determined as a function of $g_{\mu\nu}$ by defining the covariant derivative
of the metric to be zero%
\[
0=g_{\mu\nu,\rho}-g_{\mu\sigma}\Gamma_{\rho\nu}^{\sigma}-\Gamma_{\mu\rho
}^{\sigma}g_{\sigma\nu}.
\]
This gives a set of 64 equations that matches the number of independent
components of $\Gamma_{\mu\nu}^{\sigma}$ which can then be solved uniquely,
provided that the metric $g_{\mu\nu}$ is not singular. It cannot, however, be
expressed in closed form, but only perturbatively in powers of the
antisymmetric field $B_{\mu\nu}.$ There are also two possible contractions of
the curvature tensor, and therefore, unlike the real case, the action is not
unique. Both fields $G_{\mu\nu}$ and $B_{\mu\nu}$ appear explicitly in the
action, but the only symmetry present is that of diffeomorphism invariance.
Einstein did notice that this unification does not satisfy the criteria that
the field $g_{\mu\nu}$ should appear as a covariant entity with an underlying
symmetry principle. It turned out that although the field $B_{\mu\nu}$
satisfies one equation which is of the Maxwell type, the other equation
contains second order derivatives and does not imply that its antisymmetrized
field strength $\partial_{\mu}B_{\nu\rho}+\partial_{\nu}B_{\rho\mu}%
+\partial_{\rho}B_{\mu\nu}$ vanishes. In other words, the theory with
Hermitian metric on a real space-time manifold gives the interactions of the
gravitational field $G_{\mu\nu}$ and a massless field $B_{\mu\nu}.$ Much
later, it was shown that the interactions of the field $B_{\mu\nu}$ are
inconsistent at the non-linear level, because one of the degrees of freedom
becomes ghost like \cite{Damour1}. There is an exception to this in the
special case when a cosmological constant is added, in which case the theory
is rendered consistent as a mass term for the $B_{\mu\nu}$ field is acquired
\cite{Damour2},\cite{M}.

More recently, it was realized that this generalized gravity theory could be
formulated elegantly and unambiguously as a gauge theory of the $U\left(
1,3\right)  $ group \cite{chams1}. A formulation of gravity based on the gauge
principle is desirable because such an approach might give a handle on the
unification of gravity with the other interactions, all of which are based on
gauge theories. This can be achieved by taking the gauge field $\omega_{\mu
a}^{\quad b}$ to be anti-Hermitian:%
\[
\overline{\omega_{\mu a}^{\quad b}}=-\eta_{c}^{a}\omega_{\mu d}^{\quad c}%
\eta_{b}^{d},
\]
where
\[
\eta_{b}^{a}=\text{diag}\left(  -1,1,1,1\right)  ,
\]
is the Minkowski metric. A complex vielbein $e_{\mu}^{a}$ is then introduced
which transforms in the fundamental representation of the group $U(1,3).$ The
complex conjugate of $e_{\mu}^{a}$ is defined by $e_{\mu a}=\overline{e_{\mu
}^{a}}.$ The curvature associated with the gauge field $\omega_{\mu a}^{\quad
b}$ is given by%
\[
R_{\mu\nu a}^{\hspace{0.2in}b}=\partial_{\mu}\omega_{\nu a}^{\quad b}%
-\partial_{\nu}\omega_{\mu a}^{\quad b}+\omega_{\mu a}^{\quad c}\omega_{\nu
c}^{\quad b}-\omega_{\nu a}^{\quad c}\omega_{\mu c}^{\quad b}.
\]
The gauge invariant Hermitian action is uniquely given by
\[
I=%
{\displaystyle\int}
d^{4}x\left\vert e\right\vert e^{\mu a}R_{\mu\nu a}^{\hspace{0.2in}b}%
e_{b}^{\nu},
\]
where
\[
\left\vert e\right\vert ^{2}=\left(  \det e_{\mu}^{a}\right)  \left(  \det
e_{\nu a}\right)
\]
and the inverse vierbein is defined by
\[
e_{a}^{\mu}e_{\nu}^{a}=\delta_{\nu}^{\mu},\qquad e^{\mu a}=\overline
{e_{a}^{\mu}}\,.
\]
This action coincides, in the linearized approximation, with the action
proposed by Einstein, but is not identical. The reason is that in going from
first order formalism where the field $\omega_{\mu a}^{\quad b}$ is taken as
an independent field determined by its equations of motion, one gets a
non-linear equation which can only be solved perturbatively. A similar
situation is met in the Einstein theory where the solution of the metricity
condition determines the connection $\Gamma_{\mu\nu}^{\rho}$ as function of
the Hermitian metric $g_{\mu\nu}$ in a perturbative expansion. The gauge field
$\omega_{\mu a}^{\quad a}$ associated with the $U(1)$ subgroup of $U(1,3)$
couples only linearly, so that its equation of motion simplifies to
\[
\frac{1}{\sqrt{G}}\partial_{\nu}\left(  \sqrt{G}\left(  e_{a}^{\nu}e^{\mu
a}-e_{a}^{\mu}e^{\nu a}\right)  \right)  =0,
\]
where $G=\det G_{\mu\nu}.$ In the linearized approximation, this equation
takes the form $\partial_{\nu}B^{\mu\nu}=0$ which was the original motivation
for Einstein to identify $B_{\mu\nu}$with the Maxwell field \cite{Ein1}%
,\cite{Einstein}. In the gauge formulation the metric arises as a product of
the vierbeins $g_{\mu\nu}=e_{\mu}^{a}e_{\nu a}$ which satisfies the
hermiticity condition $\overline{g_{\mu\nu}}=g_{\nu\mu}$ as can be easily
verified. Decomposing the vierbein into its real and imaginary parts
\[
e_{\mu}^{a}=e_{\mu}^{a0}+ie_{\mu}^{a1},
\]
and similarly for the anti-Hermitian infinitesimal gauge parameters
\[
\Lambda_{a}^{\;b}=\Lambda_{a}^{\;b0}+i\Lambda_{a}^{\;b1},
\]
where $\Lambda_{ab}^{\quad0}=-\Lambda_{ba}^{\quad0}$ and $\Lambda_{ab}%
^{\quad1}=\Lambda_{ba}^{\quad1}.$ From the gauge transformations
\[
\delta e_{\mu}^{a}=\Lambda_{b}^{\;a}e_{\mu}^{b},
\]
we see that there exists a gauge where the antisymmetric part of $e_{\mu}%
^{a0}$ and the symmetric part of $e_{\mu}^{a1}$ can be set to zero. This shows
that the gauge theory with complex vierbeins is equivalent to the theory with
a symmetric metric $G_{\mu\nu}$ and antisymmetric field $B_{\mu\nu}.$

It turns out that the field $B_{\mu\nu}$ does not have the correct properties
to represent the electromagnetic field. Moreover, as noted by Einstein, the
fields $G_{\mu\nu}$ and $B_{\mu\nu}$ are not unified with respect to a higher
symmetry because they appear as independent tensors with respect to general
coordinate transformations. In the massless spectrum of string theory the
three fields $G_{\mu\nu}$, $B_{\mu\nu}$ and the dilaton $\phi$ are always
present. The effective action of closed string theory contains, besides the
Einstein term for the metric $G_{\mu\nu}$, a kinetic term for the field
$B_{\mu\nu}$ such that the later appears only through its field strength
\[
H_{\mu\nu\rho}=\partial_{\mu}B_{\nu\rho}+\partial_{\nu}B_{\rho\mu}%
+\partial_{\rho}B_{\mu\nu}.
\]
This implies that there is a hidden symmetry
\[
\delta B_{\mu\nu}=\partial_{\mu}\Lambda_{\nu}-\partial_{\nu}\Lambda_{\mu}%
\]
\ preventing the explicit appearance of the field $B_{\mu\nu}.$ As both
$G_{\mu\nu}$ and $B_{\mu\nu}$ fields are unified in the Hermitian field
$g_{\mu\nu}$, it will be necessary to combine the diffeomorphism parameter
$\zeta^{\mu}\left(  x\right)  $ and the abelian parameters $\Lambda_{\mu
}\left(  x\right)  $ into one complex parameter. This leads us to consider the
idea that the manifold of space-time is complex, but in such a way that at low
energies the imaginary parts of the coordinates should be very small compared
with the real ones, and become relevant only at energies near the Planck
scale. Indeed this idea was first put forward by Witten \cite{Witten} in his
study of topological orbifolds. He was motivated by the observation that
string scattering amplitudes at Planckian energies depend on the imaginary
parts of the string coordinates \cite{GM}.

We shall not require the sigma model to be topological. Instead we shall start
with the sigma model \cite{HW}, \cite{Witten2}
\[
I=%
{\displaystyle\int}
d\sigma^{+}d\sigma^{-}g_{\mu\overline{\nu}}\left(  Z(\sigma,\overline{\sigma
}),\overline{Z}\left(  \sigma,\overline{\sigma}\right)  \right)  \partial
_{+}Z^{\mu}\partial_{-}Z^{\overline{\nu}},
\]
where we have denoted the complex coordinates by $Z^{\mu}$, $\mu=1,\cdots,d$
,\ and their complex conjugates by $\overline{Z^{\mu}}\equiv Z^{\overline{\mu
}},$ and where the world-sheet coordinates are denoted by $\sigma^{\pm}%
=\sigma^{0}\pm\sigma^{1}$. We also require that the background metric for the
complex $d$-dimensional manifold M to be Hermitian so that
\[
\overline{g_{\mu\overline{\nu}}}=g_{\nu\overline{\mu}},\qquad g_{\mu\nu
}=g_{\overline{\mu}\,\overline{\nu}}=0.
\]
Decomposing the metric into real and imaginary components%
\[
g_{\mu\overline{\nu}}=G_{\mu\nu}+iB_{\mu\nu},
\]
the hermiticity condition implies that $G_{\mu\nu}$ is symmetric and
$B_{\mu\nu}$ is antisymmetric. This sigma model can be made topological by
including additional fields, but this will not be considered here. It can be
embedded into a $2d$ dimensional real sigma model with coordinates of the
target manifold denoted by $Z^{i}=\left\{  Z^{\mu},Z^{\overline{\mu}}\right\}
$, $\mu=1,\cdots,d,$ \ with a background metric $g_{ij}\left(  Z\right)  $ and
antisymmetric tensor $b_{ij}\left(  Z\right)  $, with the action
\[
I=%
{\displaystyle\int}
d\sigma^{+}d\sigma^{-}\left(  g_{ij}\left(  Z\right)  +b_{ij}\left(  Z\right)
\right)  \partial_{+}Z^{i}\partial_{-}Z^{j}.
\]
The connection is taken to be
\begin{align*}
\Gamma_{ij}^{k}  &  =\mathring{\Gamma}_{ij}^{k}+\frac{1}{2}g^{kl}T_{ijl},\\
\mathring{\Gamma}_{ij}^{k}  &  =\frac{1}{2}g^{kl}\left(  \partial_{i}%
g_{lj}+\partial_{j}g_{il}-\partial_{l}g_{ij}\right)  ,\\
T_{ijk}  &  =\left(  \partial_{i}b_{jk}+\partial_{j}b_{ki}+\partial_{k}%
b_{ij}\right)  ,
\end{align*}
so that the torsion on the target manifold is totally antisymmetric. The
embedding is defined by taking
\begin{align*}
g_{\mu\nu}  &  =0=g_{\overline{\mu}\,\overline{\nu}},\quad\\
b_{\mu\nu}  &  =0=b_{\overline{\mu}\,\overline{\nu}},\qquad\\
b_{\overline{\nu}\mu}  &  =g_{\mu\overline{\nu}}=-b_{\mu\overline{\nu}%
}=g_{\overline{\nu}\mu}%
\end{align*}
so that, as can be easily verified, the only non-zero components of the
connections are
\[
\Gamma_{\mu\lambda}^{\rho}=g^{\overline{\nu}\rho}\partial_{\lambda}%
g_{\mu\overline{\nu}}%
\]
and their complex conjugates.

Having made the identification of how the complex $d$-dimensional target
manifold is embedded into the sigma model with a $2d$ real target manifold, we
can proceed to summarize the geometrical properties of Hermitian
non-K\"{a}hler manifolds.

The Hermitian manifold $M$ of complex dimensions $d$ \ is defined as a
Riemannian manifold with real dimensions $2d$ with Riemannian metric $g_{ij}$
and complex coordinates $z^{i}=\left\{  z^{\mu},z^{\overline{\mu}}\right\}  $
where Latin indices $i,j,k,\cdots,$ run over the range $1,2,\cdots
,d,\overline{1},\overline{2},\cdots,\overline{d}.$ The invariant line element
is then \cite{Yano}
\[
ds^{2}=g_{ij}dz^{i}dz^{j},
\]
where the metric $g_{ij}$ is hybrid%
\[
g_{ij}=\left(
\begin{array}
[c]{cc}%
0 & g_{\mu\overline{\nu}}\\
g_{\nu\overline{\mu}} & 0
\end{array}
\right)  .
\]
It has also an integrable complex structure $J_{i}^{j}$ satisfying
\[
J_{i}^{k}J_{k}^{j}=-\delta_{i}^{j},
\]
and with a vanishing Nijenhuis tensor
\[
N_{ji}^{\hspace{0.06in}h}=J_{j}^{t}\left(  \partial_{t}J_{i}^{h}-\partial
_{i}J_{t}^{h}\right)  -J_{i}^{t}\left(  \partial_{t}J_{j}^{h}-\partial
_{j}J_{t}^{h}\right)  .
\]
Locally, the complex structure has components
\[
J_{i}^{j}=\left(
\begin{array}
[c]{cc}%
i\delta_{\mu}^{\nu} & 0\\
0 & -i\delta_{\overline{\mu}}^{\overline{\nu}}%
\end{array}
\right)  .
\]
The affine connection with torsion $\Gamma_{ij}^{h}$ is introduced so that the
following two conditions are satisfied%
\begin{align*}
\nabla_{k}g_{ij}  &  =\partial_{k}g_{ij}-\Gamma_{ik}^{h}g_{hj}-\Gamma_{jk}%
^{h}g_{ih}=0,\\
\nabla_{k}F_{i}^{j}  &  =\partial_{k}F_{i}^{j}-\Gamma_{ik}^{h}F_{h}^{j}%
+\Gamma_{hk}^{j}F_{i}^{h}=0.
\end{align*}
These conditions do not determine the affine connection uniquely and there
exists several possibilities used in the literature. We shall adopt the Chern
connection, which is the one most commonly used, . It is defined by
prescribing that the $(2d)^{2}$ linear differential forms
\[
\omega_{\;j}^{i}=\Gamma_{jk}^{i}dz^{k},
\]
be such that $\omega_{\;\nu}^{\mu}$ and $\omega_{\;\overline{\nu}}%
^{\overline{\mu}}$ are given by \cite{Goldberg}
\begin{align*}
\omega_{\;\nu}^{\mu}  &  =\Gamma_{\nu\rho}^{\mu}dz^{\rho},\\
\overline{\omega_{\;\nu}^{\mu}}  &  =\omega_{\;\overline{\nu}}^{\overline{\mu
}}=\Gamma_{\overline{\nu}\,\overline{\rho}}^{\overline{\mu}}dz^{\overline
{\rho}},
\end{align*}
with the remaining $(2d)^{2}$ forms set equal to zero. For $\omega_{\;\nu
}^{\mu}$ to have a metrical connection the differential of the metric tensor
$g$ must be given by
\[
dg_{\mu\overline{\nu}}=\omega_{\;\mu}^{\rho}g_{\rho\overline{\nu}}%
+\omega_{\;\overline{\nu}}^{\overline{\rho}}g_{\mu\overline{\rho}},
\]
from which we obtain%
\[
\partial_{\lambda}g_{\mu\overline{\nu}}dz^{\lambda}+\partial_{\overline
{\lambda}}g_{\mu\overline{\nu}}dz^{\overline{\lambda}}=\Gamma_{\mu\lambda
}^{\rho}g_{\rho\overline{\nu}}dz^{\lambda}+\Gamma_{\overline{\nu}%
\overline{\lambda}}^{\overline{\rho}}g_{\mu\overline{\rho}}dz^{\overline
{\lambda}},
\]
so that
\begin{align*}
\Gamma_{\mu\lambda}^{\rho}  &  =g^{\overline{\nu}\rho}\partial_{\lambda}%
g_{\mu\overline{\nu}},\\
\Gamma_{\overline{\nu}\overline{\lambda}}^{\overline{\rho}}  &  =g^{\overline
{\rho}\mu}\partial_{\overline{\lambda}}g_{\mu\overline{\nu}},
\end{align*}
where the inverse metric $g^{\overline{\nu}\mu}$ is defined by
\[
g^{\overline{\nu}\mu}g_{\mu\overline{\kappa}}=\delta_{\overline{\kappa}%
}^{\overline{\nu}}.
\]
Notice that the Chern connection agrees with the connection obtained above by
embedding of the non-linear sigma model with Hermitian target manifold into a
real one with double the number of dimensions.\ The condition $\nabla_{k}%
J_{i}^{j}=0$ is automatically satisfied and the connection is metric. The
torsion forms are defined by
\begin{align*}
\Theta^{\mu}  &  \equiv-\frac{1}{2}T_{\nu\rho}^{\hspace{0.1in}\mu}dz^{\nu
}\wedge dz^{\rho}\\
&  =\omega_{\;\nu}^{\mu}dz^{\nu}=-\Gamma_{\nu\rho}^{\mu}dz^{\nu}\wedge
dz^{\rho},
\end{align*}
which implies that%
\begin{align*}
T_{\nu\rho}^{\hspace{0.1in}\mu}  &  =\Gamma_{\nu\rho}^{\mu}-\Gamma_{\rho\nu
}^{\mu}\\
&  =g^{\overline{\sigma}\mu}\left(  \partial_{\rho}g_{\nu\overline{\sigma}%
}-\partial_{\nu}g_{\rho\overline{\sigma}}\right)  .
\end{align*}
The torsion form is related to the differential of the Hermitian form
\[
J=\frac{1}{2}J_{ij}dz^{i}\wedge dz^{j},
\]
where
\[
J_{ij}=J_{i}^{k}g_{kj}=-J_{ji},
\]
is antisymmetric and satisfy
\begin{align*}
J_{\mu\nu}  &  =0=J_{\overline{\mu}\,\overline{\nu}},\\
J_{\mu\overline{\nu}}  &  =ig_{\mu\overline{\nu}}=-J_{\overline{\nu}\mu},
\end{align*}
so that
\[
J=ig_{\mu\overline{\nu}}dz^{\mu}\wedge dz^{\overline{\nu}}.
\]
The differential of the two-form $J$ is then%
\[
dJ=\frac{1}{6}J_{ijk}dz^{i}\wedge dz^{j}\wedge dz^{k},
\]
so that
\[
J_{ijk}=\partial_{i}J_{jk}+\partial_{j}J_{ki}+\partial_{k}J_{ij}.
\]
The only non-vanishing components of this tensor are
\begin{align*}
J_{\mu\nu\overline{\rho}}  &  =i\left(  \partial_{\mu}g_{\nu\overline{\rho}%
}-\partial_{\nu}g_{\mu\overline{\rho}}\right)  =-iT_{\mu\nu}^{\hspace
{0.1in}\sigma}g_{\sigma\overline{\rho}}=-iT_{\mu\nu\overline{\rho}},\\
J_{\overline{\mu}\,\overline{\nu}\rho}  &  =-i\left(  \partial_{\overline{\mu
}}g_{\rho\overline{\nu}}-\partial_{\overline{\nu}}g_{\rho\overline{\mu}%
}\right)  =iT_{\overline{\mu}\,\overline{\nu}}^{\hspace{0.1in}\overline
{\sigma}}g_{\rho\overline{\sigma}}=iT_{\overline{\mu}\,\overline{\nu}\rho}.
\end{align*}
The curvature tensor of the metric connection is constructed in the usual
manner
\[
\Omega_{\;j}^{i}=d\omega_{\;j}^{i}-\omega_{\;k}^{i}\wedge\omega_{\;j}^{k},
\]
with the only non-vanishing components being $\Omega_{\;\mu}^{\nu}$ and
$\Omega_{\;\overline{\mu}}^{\overline{\nu}}.$ These are given by
\[
\Omega_{\;\mu}^{\nu}=-R_{\;\mu\kappa\lambda}^{\nu}dz^{\kappa}\wedge
dz^{\lambda}-R_{\;\mu\kappa\overline{\lambda}}^{\nu}dz^{\kappa}\wedge
dz^{\overline{\lambda}},
\]
where one can show that%
\begin{align*}
R_{\;\mu\kappa\lambda}^{\nu}  &  =0,\\
R_{\;\mu\kappa\overline{\lambda}}^{\nu}  &  =g^{\overline{\rho}\nu}%
\partial_{\kappa}\partial_{\overline{\lambda}}g_{\mu\overline{\rho}}%
+\partial_{\overline{\lambda}}g^{\overline{\rho}\nu}\partial_{\kappa}%
g_{\mu\overline{\rho}}.
\end{align*}
Transvecting the last relation with $g_{\nu\overline{\sigma}}$ we obtain
\[
-R_{\mu\overline{\sigma}\kappa\overline{\lambda}}=\partial_{\kappa}%
\partial_{\overline{\lambda}}g_{\mu\overline{\sigma}}+g_{\nu\overline{\sigma}%
}\partial_{\overline{\lambda}}g^{\overline{\rho}\nu}\partial_{\kappa}%
g_{\mu\overline{\rho}}.
\]
Therefore the only non-vanishing covariant components of the curvature tensor
are
\[
R_{\mu\overline{\nu}\kappa\overline{\lambda}},\quad R_{\mu\overline{\nu
\,}\,\overline{\kappa}\lambda},\quad R_{\overline{\mu}\nu\kappa\overline
{\lambda}},\quad R_{\overline{\mu}\nu\overline{\kappa}\lambda},
\]
which are related by%
\[
R_{\mu\overline{\nu}\kappa\overline{\lambda}}=-R_{\overline{\nu}\mu
\kappa\overline{\lambda}}=-R_{\mu\overline{\nu}\overline{\lambda}\kappa},
\]
and satisfy the first Bianchi identity \cite{Goldberg}%
\[
R_{\;\mu\kappa\overline{\lambda}}^{\nu}-R_{\;\kappa\mu\overline{\lambda}}%
^{\nu}=\nabla_{\overline{\lambda}}T_{\mu\kappa}^{\quad\nu}.
\]
The second Bianchi identity is given by%
\[
\nabla_{\rho}R_{\mu\overline{\nu}\kappa\overline{\lambda}}-\nabla_{\kappa
}R_{\mu\overline{\nu}\rho\overline{\lambda}}=R_{\mu\overline{\nu}%
\sigma\overline{\lambda}}T_{\rho\kappa}^{\quad\sigma},
\]
together with the conjugate relations. There are three possible contractions
for the curvature tensor which are called the Ricci tensors
\[
R_{\mu\overline{\nu}}=-g^{\overline{\lambda}\kappa}R_{\mu\overline{\lambda
}\kappa\overline{\nu}},\quad S_{\mu\overline{\nu}}=-g^{\overline{\lambda
}\kappa}R_{\mu\overline{\nu}\kappa\overline{\lambda}},\quad T_{\mu
\overline{\nu}}=-g^{\overline{\lambda}\kappa}R_{\kappa\overline{\lambda}%
\mu\overline{\nu}}.
\]
Upon further contraction these result in two possible curvature scalars%
\[
R=g^{\overline{\nu}\mu}R_{\mu\overline{\nu}},\quad S=g^{\overline{\nu}\mu
}S_{\mu\overline{\nu}}=g^{\overline{\nu}\mu}T_{\mu\overline{\nu}}.
\]
Note that when the torsion tensors vanishes, the manifold $M$ \ becomes
K\"{a}hler. We shall not impose the K\"{a}hler condition as we are interested
in Hermitian non-K\"{a}hlerian geometry. We note that it is also possible to
consider the Levi-Civita connection $\mathring{\Gamma}_{ij}^{k}$ and the
associated Riemann curvature $K_{kij}^{\hspace{0.13in}h}$ where
\begin{align*}
\mathring{\Gamma}_{ij}^{k}  &  =\frac{1}{2}g^{kl}\left(  \partial_{i}%
g_{lj}+\partial_{j}g_{il}-\partial_{l}g_{ij}\right)  ,\\
K_{kij}^{\hspace{0.13in}\,h}  &  =\partial_{k}\mathring{\Gamma}_{ij}%
^{h}-\partial_{i}\mathring{\Gamma}_{kj}^{h}+\mathring{\Gamma}_{kt}%
^{h}\mathring{\Gamma}_{ij}^{t}-\mathring{\Gamma}_{it}^{h}\mathring{\Gamma
}_{kj}^{t}.
\end{align*}
The relation between the Chern connection and the Levi-Civita connection is
given by%
\[
\Gamma_{ij}^{k}=\mathring{\Gamma}_{ij}^{k}+\frac{1}{2}\left(  T_{ij}%
^{\hspace{0.07in}k}-T_{\hspace{0.03in}\,ij}^{k}-T_{\hspace{0.03in}\,ji}%
^{k}\right)  .
\]
It can be immediately verified that the Levi-Civita connection of the
Hermitian manifold is identical to the one obtained from the non-linear sigma
model, but only after the identification of $b_{\mu\overline{\nu}}$ with
$-g_{\mu\overline{\nu}}.$ The Ricci tensor and curvature scalar are $K_{ij}=$
$K_{tij}^{\hspace{0.13in}t}$ and $K=g^{ij}K_{ij}.$ Moreover, it is also
possible to define $H_{kj}=K_{kji}^{\hspace{0.13in}\,t}J_{t}^{i}$ and
$H=g^{kj}H_{kj}$. The two scalar curvatures $K$ and $H$ are not independent
but related by \cite{Gaud}
\[
K-H=\mathring{\nabla}^{h}J^{ij}\mathring{\nabla}_{j}J_{ih}-\mathring{\nabla
}^{k}J_{ki}\mathring{\nabla}_{h}J^{hi}-2J^{ji}\mathring{\nabla}_{j}%
\mathring{\nabla}^{k}J_{ki}.
\]
There are also relations between curvatures of the Chern connection and those
of the Levi-Civita connection, mainly \cite{Gaud}%
\[
\frac{1}{2}K=S-\nabla^{\mu}T_{\mu}-\nabla^{\overline{\mu}}T_{\overline{\mu}%
}-T_{\mu}T_{\overline{\nu}}g^{\overline{\nu}\mu},
\]
where $T_{\mu}=T_{\mu\nu}^{\hspace{0.1in}\nu}.$ There are two natural
conditions that can be imposed on the torsion. The first is $T_{\mu}=0$ which
results in a semi-K\"{a}hler manifold. The other is when the torsion is
complex analytic so that $\nabla_{\overline{\lambda}}T_{\mu\kappa}%
^{\hspace{0.1in}\nu}=0$ implying that the curvature tensor has the same
symmetry properties as in the K\"{a}hler case. In this work we shall not
impose any conditions on the torsion tensor.

We note that the line element
\[
ds^{2}=2g_{\mu\overline{\nu}}dz^{\mu}d\overline{z}^{\overline{\nu}},
\]
preserves its form under infinitesimal holomorphic transformations%
\begin{align*}
z^{\mu}  &  \rightarrow z^{\mu}-\zeta^{\mu}\left(  z\right)  ,\\
z^{\overline{\mu}}  &  \rightarrow z^{\overline{\mu}}-\zeta^{\overline{\mu}%
}\left(  \overline{z}\right)  ,
\end{align*}
as can be seen from the transformations%
\[
\delta g_{\mu\overline{\nu}}=\partial_{\mu}\zeta^{\lambda}g_{\lambda
\overline{\nu}}+\partial_{\overline{\nu}}\zeta^{\overline{\lambda}}%
g_{\mu\overline{\lambda}}+\zeta^{\lambda}\partial_{\lambda}g_{\mu\overline
{\nu}}+\zeta^{\overline{\lambda}}\partial_{\overline{\lambda}}g_{\mu
\overline{\nu}}.
\]
It is instructive to express these transformations in terms of the fields
$G_{\mu\nu}(x,y)$ and $B_{\mu\nu}(x,y)$ by writing
\begin{align*}
\zeta^{\mu}(z)  &  =\alpha^{\mu}(x,y)+i\beta^{\mu}(x,y),\\
\zeta^{\overline{\mu}}(\overline{z})  &  =\alpha^{\mu}(x,y)-i\beta^{\mu}(x,y).
\end{align*}
The holomorphicity conditions on $\zeta^{\mu}$ and $\zeta^{\overline{\mu}}$
imply the relations
\begin{align*}
\partial_{\mu}^{y}\beta^{\nu}  &  =\partial_{\mu}^{x}\alpha^{\nu},\\
\partial_{\mu}^{y}\alpha^{\nu}  &  =-\partial_{\mu}^{x}\beta^{\nu},
\end{align*}
where we have denoted
\[
\partial_{\mu}^{y}=\frac{\partial}{\partial y^{\mu}},\qquad\partial_{\mu}%
^{x}=\frac{\partial}{\partial x^{\mu}}.
\]
The transformations of $G_{\mu\nu}(x,y)$ and $B_{\mu\nu}(x,y)$ are then given
by%
\begin{align*}
\delta G_{\mu\nu}(x,y)  &  =\partial_{\mu}^{x}\alpha^{\lambda}G_{\lambda\nu
}+\partial_{\nu}^{x}\alpha^{\lambda}G_{\mu\lambda}+\alpha^{\lambda}%
\partial_{\lambda}^{x}G_{\mu\nu}\\
&  -\partial_{\mu}^{x}\beta^{\lambda}B_{\lambda\nu}+\partial_{\nu}^{x}%
\beta^{\lambda}B_{\mu\lambda}+\beta^{\lambda}\partial_{\lambda}^{y}G_{\mu\nu
},\\
\delta B_{\mu\nu}(x,y)  &  =\partial_{\mu}^{x}\beta^{\lambda}G_{\lambda\nu
}-\partial_{\nu}^{x}\beta^{\lambda}G_{\mu\lambda}+\alpha^{\lambda}%
\partial_{\lambda}^{x}B_{\mu\nu}\\
&  +\partial_{\mu}^{x}\alpha^{\lambda}B_{\lambda\nu}+\partial_{\nu}^{x}%
\alpha^{\lambda}B_{\mu\lambda}+\beta^{\lambda}\partial_{\lambda}^{y}B_{\mu\nu
}.
\end{align*}
One readily recognizes that in the vicinity of small $y^{\mu}$ the fields
$G_{\mu\nu}(x,0)$ and $B_{\mu\nu}(x,0)$ transform as symmetric and
antisymmetric tensors with gauge parameters $\alpha^{\mu}(x)$ and $\beta^{\mu
}(x)$ where
\begin{align*}
\alpha^{\mu}(x,y)  &  =\alpha^{\mu}(x)-\partial_{\nu}^{x}\beta^{\mu}(x)y^{\nu
}+O(y^{2}),\\
\beta^{\mu}(x,y)  &  =\beta^{\mu}(x)+\partial_{\nu}^{x}\alpha^{\mu}(x)y^{\nu
}+O(y^{2}),
\end{align*}
as implied by the holomorphicity conditions. Therefore, it should be possible
to find an action where diffeomorphism invariance in the complex dimensions
imply diffeomorphism invariance in the real submanifold and abelian invariance
for the field $B_{\mu\nu}\left(  x\right)  $ to insure that the later only
appears through its field strength.

For simplicity, we shall now specialize to four complex dimensions. We start
with the most general action limited to derivatives of order two
\[
I=%
{\displaystyle\int\limits_{M^{4}}}
d^{4}zd^{4}\overline{z}g\left(  aR+bS+c\,T_{\mu\nu\overline{\kappa}%
}T_{\overline{\rho}\,\overline{\sigma}\lambda}g^{\overline{\rho}\mu
}g^{\overline{\sigma}\nu}g^{\overline{\kappa}\lambda}+d\,T_{\mu\nu
\overline{\kappa}}T_{\overline{\rho}\,\overline{\sigma}\lambda}g^{\overline
{\rho}\mu}g^{\overline{\sigma}\lambda}g^{\overline{\kappa}\nu}\right)  .
\]
One can show that by requiring the linearized action, in the limit
$y\rightarrow0,$ to give the correct kinetic terms for $G_{\mu\nu}(x)$ and
$B_{\mu\nu}(x)$ relates the coefficients $a,$ $b,$ $c,$ $d$ \ to each other
\cite{chams3}%
\[
b=-a,\qquad d=-1-a,\qquad c=\frac{1}{2}.
\]
In this case the action simplifies to the very elegant form
\[
I=-\frac{1}{2}%
{\displaystyle\int\limits_{M}}
d^{4}zd^{4}\overline{z}\epsilon^{\overline{\kappa}\overline{\lambda}%
\overline{\sigma}\,\overline{\eta}}\epsilon^{\mu\nu\rho\tau}g_{\tau
\overline{\eta}}\partial_{\mu}g_{\nu\overline{\sigma}}\partial_{\overline
{\kappa}}g_{\rho\overline{\lambda}}.
\]
which can be expressed in terms of the two-form $J$,
\[
I=\frac{i}{2}%
{\displaystyle\int\limits_{M}}
J\wedge\partial J\wedge\overline{\partial}J.
\]
We stress that this action is only invariant under holomorphic
transformations. The equations of motion are given by
\[
\epsilon^{\overline{\kappa}\overline{\lambda}\overline{\sigma}\,\overline
{\eta}}\epsilon^{\mu\nu\rho\tau}\left(  g_{\nu\overline{\sigma}}\partial_{\mu
}\partial_{\overline{\kappa}}g_{\rho\overline{\lambda}}+\frac{1}{2}%
\partial_{\mu}g_{\nu\overline{\sigma}}\partial_{\overline{\kappa}}%
g_{\rho\overline{\lambda}}\right)  =0,
\]
which are trivially satisfied when the metric $g_{\mu\overline{\nu}}$ is
K\"{a}hler%
\[
\partial_{\mu}g_{\nu\overline{\rho}}=\partial_{\nu}g_{\mu\overline{\rho}%
},\qquad\partial_{\overline{\sigma}}g_{\nu\overline{\rho}}=\partial
_{\overline{\rho}}g_{\nu\overline{\sigma}}.
\]
We proceed to evaluate the four-dimensional limit of the action when the
imaginary parts of the coordinates are small at low-energy. The action is a
function of the fields $G_{\mu\nu}\left(  x,y\right)  $ and $B_{\mu\nu}\left(
x,y\right)  $ which depend continuously on the coordinates $y^{\mu},$ implying
a continuos spectrum with an infinite number of fields depending on $x^{\mu}$
only. To obtain a discrete spectrum a certain physical assumption should be
made that forces the imaginary coordinates to be small. One idea, suggested by
Witten \cite{Witten}, is to suppress the imaginary parts by constructing an
orbifold space $M^{\prime}=M/G$ where $G$ is the group of imaginary shifts%
\[
z^{\mu}\rightarrow z^{\mu}+i(2\pi k^{\mu}),
\]
where $k^{\mu}$ are real. To maintain invariance under general coordinate
transformations we must require $k^{\mu}\left(  x,y\right)  $ to be coordinate
dependent. It is not easy, however, to deal with such an orbifold in field
theoretic considerations.

Instead, we shall proceed by examining the dynamical properties of the action
which depends on terms not higher than second derivatives of the fields. It is
then enough to expand the fields to second order in $y^{\mu}$ and take the
limit $y\rightarrow0.$ We therefore write
\begin{align*}
G_{\mu\nu}\left(  x,y\right)   &  =G_{\mu\nu}(x)+G_{\mu\nu\rho}(x)y^{\rho
}+\frac{1}{2}G_{\mu\nu\rho\sigma}\left(  x\right)  y^{\rho}y^{\sigma}%
+O(y^{3}),\\
B_{\mu\nu}\left(  x,y\right)   &  =B_{\mu\nu}(x)+B_{\mu\nu\rho}(x)y^{\rho
}+\frac{1}{2}B_{\mu\nu\rho\sigma}\left(  x\right)  y^{\rho}y^{\sigma}%
+O(y^{3}).
\end{align*}
In the absence of a symmetry principle that determines the fields $G_{\mu
\nu\rho}(x)$, $B_{\mu\nu\rho}(x),G_{\mu\nu\rho\sigma}(x)$ and $B_{\mu\nu
\rho\sigma}(x)$ and all higher terms as functions of $G_{\mu\nu}(x)$,
$B_{\mu\nu}(x)$ we impose boundary conditions, in the limit $y\rightarrow0,$
on the first and second derivatives of the Hermitian metric. In order to have
this action identified with the string effective action, the equations of
motion in the $y\rightarrow0$ limit should reproduce the low-energy limit of
the string equations%
\begin{align*}
0  &  =G^{\eta\tau}\left(  R\left(  G\right)  +\frac{1}{6}H_{\mu\nu\rho}%
H^{\mu\nu\rho}\right)  -2\left(  R^{\eta\tau}\left(  G\right)  +\frac{1}%
{4}H_{\hspace{0.04in}\nu\rho}^{\eta}H^{\tau\nu\rho}\right)  ,\\
0  &  =\nabla^{\mu\left(  G\right)  }H_{\mu\eta\tau}.
\end{align*}
These equations could be derived from the equations of motion of the Hermitian
theory, provided we impose the following boundary conditions on torsion and
curvature of the Hermitian manifold:
\begin{align*}
T_{\mu\nu\overline{\rho}}|_{y\rightarrow0}  &  =2iB_{\mu\nu,\rho}\left(
x\right)  ,\\
\left[  R_{\mu\overline{\sigma}\kappa\overline{\lambda}}-R_{\kappa
\overline{\sigma}\mu\overline{\lambda}}\right]  _{y\rightarrow0}  &
=-2\left(  R_{\mu\kappa\sigma\lambda}\left(  G\right)  +i\left(
\nabla_{\lambda}^{G}H_{\mu\kappa\sigma}-\nabla_{\sigma}^{G}H_{\mu\kappa
\lambda}\right)  \right)  .
\end{align*}
The solution of the torsion constraint gives, to lowest orders,
\begin{align*}
G_{\mu\nu\rho}\left(  x\right)   &  =\partial_{\nu}B_{\mu\rho}\left(
x\right)  +\partial_{\mu}B_{\nu\rho}\left(  x\right)  ,\\
B_{\mu\nu\rho}\left(  x\right)   &  =-G_{\mu\rho,\nu}\left(  x\right)
+G_{\nu\rho,\mu}\left(  x\right)  ,
\end{align*}
where all derivatives are now with respect to $x^{\mu}.$ Substituting these
into the curvature constraints yield%
\begin{align*}
G_{\mu\sigma\kappa\lambda}\left(  x\right)   &  =\partial_{\sigma}%
\partial_{\lambda}G_{\mu\kappa}\left(  x\right)  +\partial_{\mu}%
\partial_{\lambda}G_{\sigma\kappa}\left(  x\right)  +\partial_{\sigma}%
\partial_{\kappa}G_{\mu\lambda}\left(  x\right) \\
&  +\partial_{\mu}\partial_{\kappa}G_{\sigma\lambda}\left(  x\right)
-\partial_{\kappa}\partial_{\lambda}G_{\mu\sigma}\left(  x\right)  +O\left(
\partial G,\partial B\right)  ,\\
B_{\mu\sigma\kappa\lambda}\left(  x\right)   &  =\partial_{\sigma}%
\partial_{\lambda}B_{\mu\kappa}\left(  x\right)  -\partial_{\mu}%
\partial_{\lambda}B_{\sigma\kappa}\left(  x\right)  +\partial_{\sigma}%
\partial_{\kappa}B_{\mu\lambda}\left(  x\right) \\
&  -\partial_{\mu}\partial_{\kappa}B_{\sigma\lambda}\left(  x\right)
-\partial_{\kappa}\partial_{\lambda}B_{\mu\sigma}\left(  x\right)  +O\left(
\partial G,\partial B\right)  ,
\end{align*}
where $O\left(  \partial G,\partial B\right)  $ are terms of second order
\cite{chams3}.

This is encouraging, but more work is needed to establish the exact connection
between string theory effective actions and gravity on Hermitian manifolds and
not only to second order. For this to happen, one must determine,
unambiguously, the symmetry principle that restricts the continuous spectrum
as function of the imaginary coordinates to a discrete one.

To summarize, the idea that complex dimensions play a role in physics is quite
old \cite{Synge}. So far it has provided a technical advantage in obtaining
extensions and new solutions to the Einstein equations, or in providing
elegant formulations of some field theories such as Yang-Mills theory in terms
of twister spaces. At present, there is only circumstantial evidence, coming
from the study of high-energy behavior of string scattering amplitudes, where
it was observed that the imaginary parts of the string coordinates of the
target manifold appear. The work presented here is an attempt to show that it
might be possible to formulate geometrically the effective string theory for
target manifolds with complex dimensions. In this picture the metric tensor
and antisymmetric tensor of the effective theory are unified in one field, the
metric tensor of the Hermitian manifold, an idea first put forward by Einstein.

\bigskip{\huge Acknowledgments\medskip}

I\ would like to thank Dr. Ismail Serageldin, Director of Bibliotheca
Alexandrina, and Professor Edward Witten for their kind invitation to the
Einstein Symposium 2005, which was an extraordinary and stimulating event.
This research is supported in part by the National Science Foundation under
Grant No. Phys-0313416.

\end{document}